\documentclass[
superscriptaddress,
twocolumn,
showpacs,
bibnotes,
amsmath,
amssymb,
aps,
prb,
floatfix,
]{revtex4-2}

\usepackage{graphicx}
\usepackage{dcolumn}
\usepackage{bm}
\usepackage{cleveref}
\usepackage{silence}
\usepackage{ulem}
\usepackage{color}
\usepackage{soul}

\begin{document}


\title{$c$-axis Transport in UTe$_{2}$: Evidence of Three Dimensional Conductivity Component} 

\author{Yun Suk Eo}
    \affiliation{Maryland Quantum Materials Center and Department of Physics, University of Maryland, College Park, Maryland 20742, USA}
\author{Shouzheng Liu}
    \affiliation{Department of Physics, New York University, New York, New York, USA}    
\author{Shanta R. Saha}
    \affiliation{Maryland Quantum Materials Center and Department of Physics, University of Maryland, College Park, Maryland 20742, USA}
\author{Hyunsoo Kim}
    \altaffiliation{Current address: Department of Physics, Missouri University of Science and Technology, Rolla, Missouri 65409, USA}
    \affiliation{Maryland Quantum Materials Center and Department of Physics, University of Maryland, College Park, Maryland 20742, USA}
\author{Sheng Ran}
     \altaffiliation{Current address: Department of Physics, Washington University in St. Louis, St. Louis, Missouri, 63130, USA}
     \affiliation{Maryland Quantum Materials Center and Department of Physics, University of Maryland, College Park, Maryland 20742, USA}
     \affiliation{NIST Center for Neutron Research, National Institute of Standards and Technology, Gaithersburg, Maryland 20899, USA}
\author{Jarryd A. Horn}
    \affiliation{Maryland Quantum Materials Center and Department of Physics, University of Maryland, College Park, Maryland 20742, USA}
\author{Halyna Hodovanets}
    \altaffiliation{Current address: Department of Physics, Missouri University of Science and Technology, Rolla, Missouri 65409, USA}
    \affiliation{Maryland Quantum Materials Center and Department of Physics, University of Maryland, College Park, Maryland 20742, USA}
\author{John Collini}
    \affiliation{Maryland Quantum Materials Center and Department of Physics, University of Maryland, College Park, Maryland 20742, USA}
\author{Tristin Metz}
    \affiliation{Maryland Quantum Materials Center and Department of Physics, University of Maryland, College Park, Maryland 20742, USA}
\author{Wesley T. Fuhrman}
    \affiliation{Maryland Quantum Materials Center and Department of Physics, University of Maryland, College Park, Maryland 20742, USA}
\author{Andriy~H.~Nevidomskyy}
    \affiliation{Department of Physics and Astronomy, Rice University, Houston, Texas 77005, USA}
\author{Jonathan D. Denlinger}
    \affiliation{Advanced Light Source, Lawrence Berkeley National Laboratory, Berkeley, CA 94720, USA}
\author{Nicholas P. Butch}
    \affiliation{Maryland Quantum Materials Center and Department of Physics, University of Maryland, College Park, Maryland 20742, USA}
     \affiliation{NIST Center for Neutron Research, National Institute of Standards and Technology, Gaithersburg, MD 20899, USA}
\author{Michael S. Fuhrer}
    \affiliation{Monash University, Melbourne, Victoria 3800, Australia}
    \affiliation{ARC Centre of Excellence in Future Low-Energy Electronics Technologies, Monash University, Victoria 3800 Australia}
\author{L. Andrew Wray}
    \affiliation{Department of Physics, New York University, New York, New York, USA}        
\author{Johnpierre Paglione}
    \affiliation{Maryland Quantum Materials Center and Department of Physics, University of Maryland, College Park, Maryland 20742, USA}
    \affiliation{Canadian Institute for Advanced Research, Toronto, Ontario, Canada M5G 1Z8}
    \email{paglione@umd.edu}
    
\date{\today}
    
\begin{abstract}
We study the temperature dependence of electrical resistivity for currents directed along all crystallographic axes of the spin-triplet superconductor UTe$_{2}$. We focus particularly on an accurate determination of the resistivity along the $c$-axis ($\rho_c$) by using a generalized Montgomery technique that allows extraction of  crystallographic resistivity components from a single sample. In contrast to expectations from the observed highly anisotropic band structure, our measurement of the absolute values of resistivities in all current directions reveals a surprisingly nearly isotropic transport behavior at temperatures above Kondo coherence, with $\rho_c \sim \rho_b \sim 2\rho_a$, that evolves to reveal qualitatively distinct behaviors on cooling. The temperature dependence of $\rho_c$ exhibits a peak at a temperature much lower than the onset of Kondo coherence observed in $\rho_a$ and $\rho_b$, consistent with features in magnetotransport and magnetization that point to a magnetic origin.  A comparison to the temperature-dependent evolution of the scattering rate observed in angle-resolved photoemission spectroscopy experiments provides important insights into the underlying electronic structure necessary for building a microscopic model of superconductivity in UTe$_{2}$. 

\end{abstract}


\maketitle

 
The recently discovered superconductivity in UTe$_{2}$ \cite{ran2019nearly} is believed to be a strong contender for spin-triplet Cooper pairing driven by ferromagnetic spin fluctuations, as suggested by scaling of magnetization data \cite{ran2019nearly}, muon spin relaxation experiments \cite{sundar2019coexistence}, and an upper critical field that greatly exceeds the Pauli paramagnetic limit along all principal axes \cite{ran2019nearly}. A point-nodal structure in the superconducting gap is evidenced by studies of thermal conductivity and penetration depth \cite{metz2019point}, and the temperature dependence of the Knight shift in nuclear magnetic resonance is weak, which is consistent with the degeneracy existing in the spin-triplet state \cite{ran2019nearly, nakamine2019superconducting}. 
Other fascinating properties including re-entrant superconductivity  \cite{ran2019extreme, knebel2019field} and pressure-induced multiple superconducting phases \cite{braithwaite2019multiple, ran2020enhancement} signal a rich superconducting state in UTe$_{2}$. Observations of a split-transition in thermodynamic critical temperature ($T_{c}$) at ambient pressure and the existence of the Kerr effect at $T_{c}$, indicating breaking of time-reversal symmetry in the superconducting state, point to a two-component order parameter, expected in a topological Weyl superconductor \cite{hayes2020weyl, nevidomskyy2020two-component}.
Together with observations of novel surface states \cite{jiao2020chiral,bae2019anomalous}, magnetic excitation spectra \cite{PRL_Duan, knafo2021low,duan2021resonance,butch2022symmetry}, and tunability of the transition temperature and splitting \cite{thomas2021spatially,rosa2022single},  the plethora of interesting phenomena in UTe$_{2}$ will require continued attention to the details of this fascinating system\cite{aoki2022REview}.

\begin{figure}
    \centering
    \includegraphics[scale=0.4]{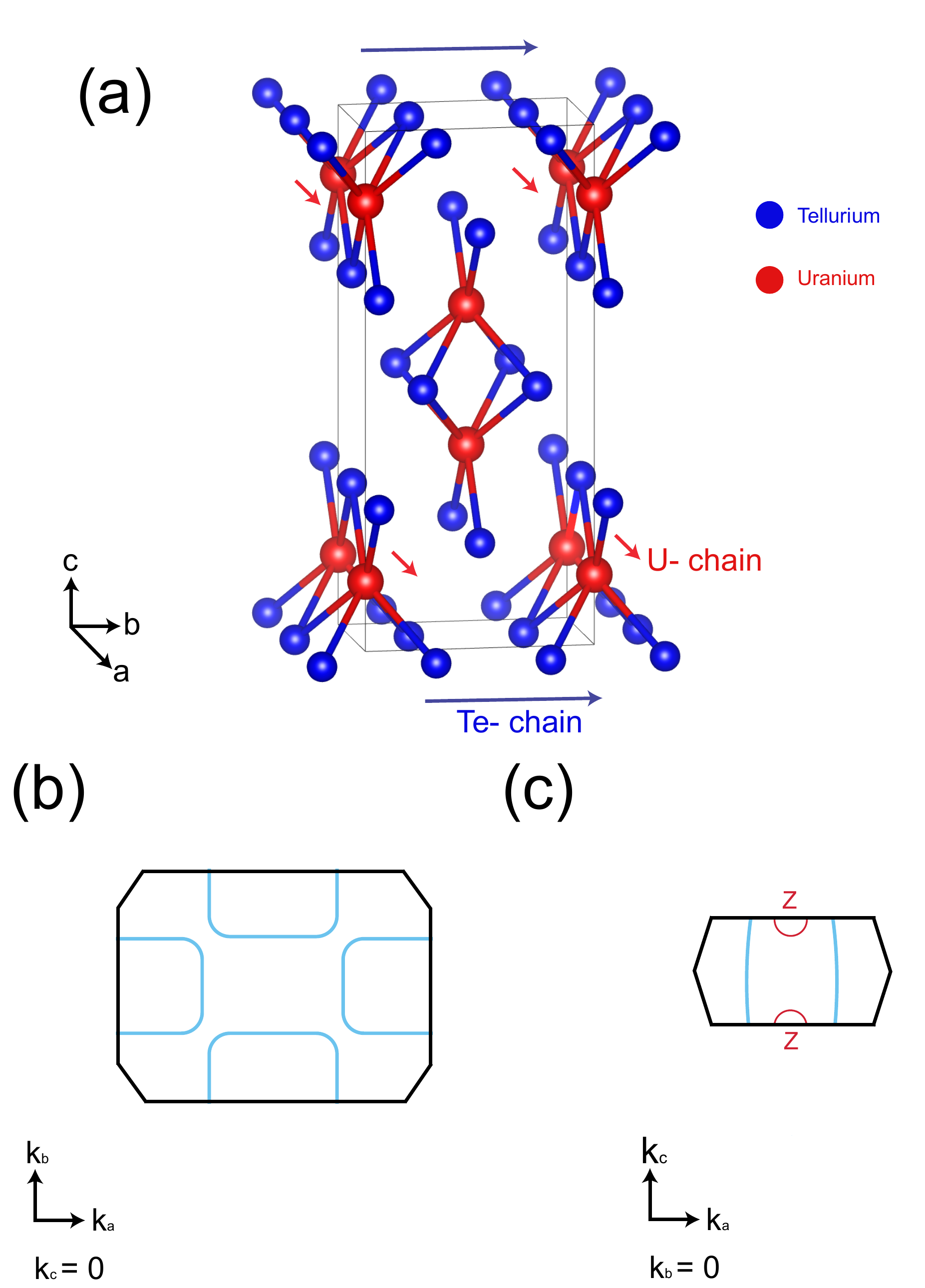}
    \caption{ (Color) UTe$_2(a)$. Crystal structure and Fermi surface. (a) Crystal structure of UTe$_2$ $a = 4.161\AA$, $b = 6.122\AA$, $c=13.955\AA$ (b) Schematic picture of the rectangular Fermi pockets (shown in blue) in the $a-b$ plane of the BZ (Based on Ref.~\cite{miao2020low}). (c) Schematic picture of the $Z$-pocket in the presence of the less dispersive rectangular pocket in $a-c$ plane of the BZ (Based on Ref.~\cite{miao2020low}).}
    \label{fig:UTe2_overview}
\end{figure}

To date, the majority of experiments have focused on elucidating the symmetry and topological class of the superconducting order parameter, or probing the landscape of proximate ground states, such as magnetism \cite{braithwaite2019multiple,lin2020phase,ran2020enhancement, niu2020fermi,niu2020evidence,thomas2020evidence}.
However, how the Fermi surface forms by the chains of uranium and tellurium atoms along the a- and $b$-axes, respectively, as shown in Fig.~(\ref{fig:UTe2_overview})~(a), together with Kondo physics and $f$-electron contributions remains an open question. 
Band calculations seem to depend sensitively on the on-site Coulomb interaction strength ($U_{\textrm{int}}$) and the role of $f$-electron physics. Local density approximation (LDA) calculations suggested that the normal state of UTe$_{2}$ is a semimetal \cite{aoki2019unconventional, fujimori2019electronic}, while more recent LDA+$U$ calculations find that a insulator-to-metal evolution can be tuned by the strength of $U_{\textrm{int}}$, with two perpendicular Fermi surface (FS) sheets forming a quasi-two-dimensional (2D) FS emerging when $U_{\textrm{int}}$ is tuned to $\sim 2$~eV \cite{shick2019spin,ishizuka2019insulator}.
Recent angle-resolved photoemission spectroscopy (ARPES) experiments at 20~K indeed observed this 2D FS in addition to a more three-dimensional (3D) $f$-like pocket surrounding the $Z$-point ($Z$ pocket) \cite{miao2020low}, as shown schematically in Fig.~(\ref{fig:UTe2_overview})~(b) and ~(c). Importantly, and without the need to invoke $U_{\textrm{int}}$, density functional theory combined with dynamical mean-field theory (DFT + DMFT) band calculations in the same study suggest the two sets of sheets comprising the quasi-2D FS derive from the U-6$d$ and Te-5$p$ orbitals of the two perpendicular chains of uranium and tellurium atoms (Fig.~(\ref{fig:UTe2_overview})~(a)), but they fail to predict the existence of the $f$-like $Z$ pocket, leaving the role of 5$f$ electrons unanswered. 

Given the confluence of interaction- and dimension-dependent contributions to the normal state electronic behavior in UTe$_{2}$, it is imperative to have an accurate measure and understanding of the conductivity anisotropy in this system in order to understand the Fermiology that leads to pairing.
Here we accurately determine the electrical resistivity along all primary crystallographic directions in UTe$_{2}$, focusing on the so-far elusive $c$-axis transport behavior in order to help elucidate the role of dimensionality and orbital contributions to the normal state electronics.
We compare the measured transport anisotropy and its temperature dependences with ARPES in order to better connect peculiar behaviors with specific band components, providing a consistent picture of transport in UTe$_{2}$. Furthermore, our magnetotransport analysis suggests magnetism as a potential origin of the qualitatively anisotropic scattering behavior at low temperatures.

\begin{figure}[th]
\begin{center}
\includegraphics[scale=1.2]{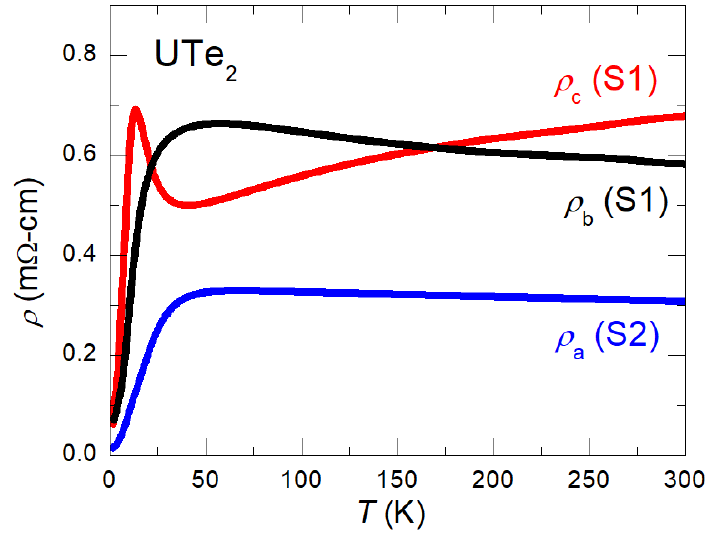}
\caption[]{(Color) Electrical resistivity of UTe$_2$ extracted using a generalized Montgomery measurement technique on two crystalline samples, including a diamond-shaped sample with $b-c$ plane orientation (sample S1) and a nearly rectangular shaped sample with $a-c$ plane orientation (sample S2). Absolute resistivities are obtained by extracting principal components of resistivities from a combination of resistance measurement geometries and numerical modeling (see SM \cite{SM} for more details, including extracted $\rho_c$ data for sample S2 and sample photos in Fig.~S6).} 
\label{Fig:resistivityresults}
\end{center}
\end{figure}

Although it is common to study transport anisotropy using the Montgomery technique\cite{montgomery1971method}, which allows extraction of two components of the resistivity tensor from a single rectangular-shaped sample, in a highly anisotropic system the possible misalignment between the sample geometry edge and crystal axis can lead to spurious results, mixing low and high conductivity channels that introduce large errors when converting to resistivity. 
We utilize a generalized (i.e. non-rectangular) Montgomery technique, where electrical contacts are placed on corners of a sample with currents directed along a mixture of principal axis directions, and employ finite element analysis to extract the principal components. We present data from a diamond-shaped sample with $b-c$ plane orientation (sample S1) and a rectangular-shaped sample with $a-c$ plane orientation (sample S2). (Details of the transport setup and considerations, sample geometries, and detailed analysis are found in  Supplemental Materials (SM) \cite{SM} sections I and II.) 
By comparing the $c$-axis components measured in both samples, we obtain an accurate absolute measurement of the $c$-axis resistivity and rule out the possibility of misinterpreting its magnitude, which has been a known issue in other quasi-2D materials\cite{Tanatar_ironpnictide}.

\begin{figure*}[th]
    \centering
    \includegraphics[scale=0.5]{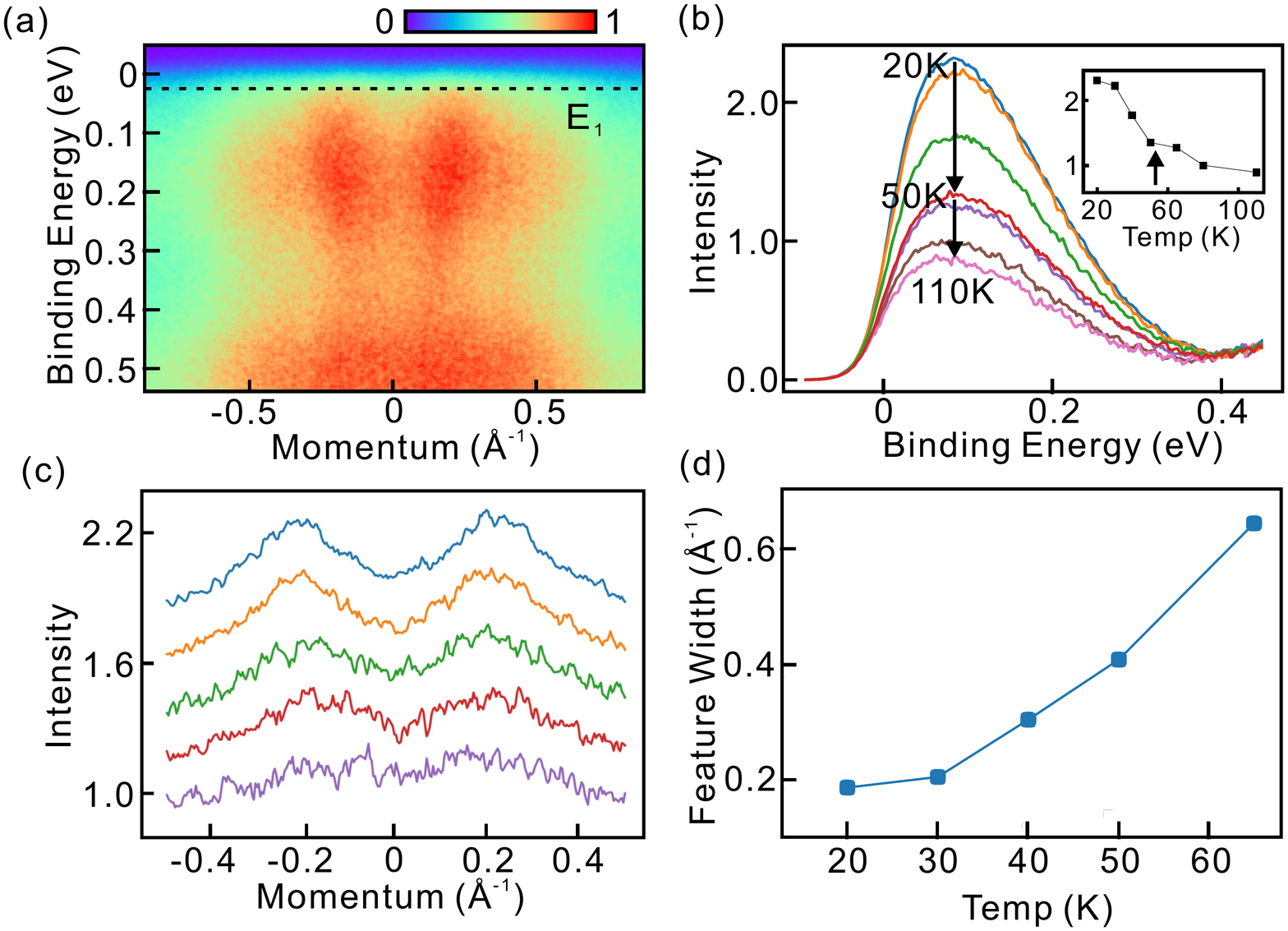}
    \caption{Temperature dependence of ARPES. (a) An ARPES image of the UTe$_2$ 6$d$ bands, measured at 20 K along the $\Gamma-X$ axis at $h\nu$ =74 eV, in normal emission from the [011] crystal face. (b) Temperature dependence of quasi-elastically scattered photoelectrons. ARPES intensity on the uranium O-edge resonance ($h\nu$= 110 eV) was integrated in a region with no visible bands (from $k=0.6 (\AA^{-1})$ to $1.0 (\AA^{-1})$). (c) Momentum distribution curves (MDC) of 6$d$ band electrons at 25~meV, measured at $h\nu$=74 eV and at temperatures of 20, 30, 40, 50, at 65 K, from top to bottom. (d) The feature width from Lorentzian fits (see \cite{SM} Sect. VIII, for details) of the MDCs in panel (c), used for comparison to resistivity (see text).}
    \label{fig:ARPES}
\end{figure*}

Figure~(\ref{Fig:resistivityresults}) presents the extracted resistivities for all three primary crystal directions, allowing analysis of the quantitative anisotropy. Our results are qualitatively consistent with the previous studies reporting $\rho_a$ and $\rho_b$, but quantitatively different by up to a factor of $\sim 2$~\cite{ran2019nearly, aoki2019unconventional}. 
In contrast to the naive expectations for the quasi-2D Fermi surfaces of UTe$_2$, the nearly isotropic conductivities 
as observed in the highly anisotropic metal in the normal state of Sr$_2$RuO$_4$\cite{NormalSr2RuO4}. 
can only be explained by the presence of a much more isotropic Fermi surface component. Here we employ a simple two channel Drude model as a start, finding quantitative agreement with the available ARPES data \cite{miao2020low}. In this model, we assume that the conductivity consists of two conduction channels, one 2D and one 3D, corresponding to the U-6$d$ and Te-5$p$ derived FS sheets and the isotropic highly U-$f$-weighted $Z$-pocket, respectively, as depicted in  Fig.~(\ref{fig:UTe2_overview}). The conductivity along the a-axis is composed of two contributions:
\begin{equation}
    \sigma_{ab} = \sigma_{2D} + \sigma_{Z}.
\end{equation}    
Since the rectangular pockets are weakly dispersive in the $c$-axis direction, we ignore their contribution and only consider the $Z$-pocket, i.e. $\sigma_{c} \approx \sigma_{Z}$. Using the Drude picture for transport, we can compare the transport data with ARPES data using an equation for the contribution of the 2D-like rectangular pocket along the $a$-axis direction, 
\begin{equation}
     \sigma_{2D} \approx 1/\rho_{a} - 1/\rho_{c} = 2n_{2D}e \mu_{2D},
\label{eq:DrudeARPES}
\end{equation}

where $n_{2D},~ m_{2D}$, and $\mu_{2D}$ refer to the carrier density, effective mass, and mobility of the 2D Fermi surfaces, respectively. The factor of two originates from the fact that two rectangular pockets exist in the Brillouin zone (BZ). We estimate the ARPES parameters (right side of Eq.~(\ref{eq:DrudeARPES})) from the uranium 6$d$ band dispersion, which predominantly contributes to the transport along the (U chain) $a$-axis (See \cite{SM} Sect. VII for details). By comparing these two experiments, we estimate a mean free path  $l_{2D}\approx 19~\AA$ at 20~K, or a mobility of 1.8 (cm$^2$/V-sec). Using the momentum distribution curves (MDCs) from ARPES at 20~K (see Fig.~(\ref{fig:ARPES})~(d)), the mobility is 2.3 (cm$^2$/V-sec), in excellent agreement. We will discuss the temperature evolution of the MDC below.  

Continuing the analysis, the Z-pocket mobility is 4.3 (cm$^2$/V-sec) at 20~K. At lower temperatures, by extrapolating the $T^2$ behavior to the zero temperature limit, we find an improvement of mobility of 29.1 (cm$^2$/V-sec) and 26.9 (cm$^2$/V-sec), for the 2D-like Fermi surface and the Z-pocket, respectively. 
We note that this two channel model does not capture differences between the $a$- and $b$-axis resistivities since we have assumed the quasi-2D channel is isotropic in the $ab$ plane. Further corrections to the two channel model, capturing this anisotropy difference, can be made by adding corrugations of the 2D-Fermi surface along the $c$-axis direction or the anisotropy of the Z-pocket. We await future ARPES studies estimating the anisotropy of the Z-pocket in all three directions and the quasiparticle lifetime along the Te-chain $b$-direction to resolve this.

Next, we discuss the temperature dependence of the resistivities, focusing on three regimes. 

Although resistivities for all three axes undergo a large drop upon cooling to low temperatures,
there is a qualitatively distinct temperature dependence between $c$-axis transport and that in the $a$-$b$ plane. 
As shown in Fig.~\ref{Fig:resistivityresults}, the resistivities start from a relatively high magnitude and drop rapidly below $\sim$ 50 K or less, with a qualitative difference found in $\rho_c$ which drops at lower temperatures than the other two components. Particularly in the intermediate temperature regime, where the behavior of $\rho(T)$ is richest, we compare with the temperature-dependent ARPES data. 

\textit{High temperature} ($50\; \text{K} \lesssim  T \lesssim 300$~K): 
At high temperatures, the most apparent contrast in resistivity behavior is in the metallic-like ($d\rho/dT <0$) vs. non-metallic-like ($d\rho/dT >0$) behavior of $\rho_c$ vs. $\rho_a$ and $\rho_b$, respectively. The weak increase of $\rho_a$ and $\rho_b$ upon cooling is attributed to single-ion Kondo behavior preceding the development of lattice coherence (although extracting a Kondo temperature is problematic due to its weak behavior, as detailed in SM \cite{SM} Section IV).
In contrast, $\rho_c$ instead exhibits a metallic-like decrease on cooling. While definitely not Kondo-like, its weak temperature dependence also suggests that it is not simply a linear behavior due to electron-phonon scattering, suggesting that a single scattering mechanism may not be dominating. 
We note also that all three resistivities in this temperature window are larger than 0.3~m$\Omega$cm, which for typical metals is approaching the Anderson localization regime \cite{imry2002introduction} as well as the Ioffe-Regel criterion for a highly anisotropic system (See SM \cite{SM} Sect. IV. for more detail), but point to the lack of any obvious hopping conduction to rule out this scenario.

\textit{Intermediate temperature} ($5\;\text{K} \lesssim T \lesssim 50$~K): 
In the intermediate-temperature regime, the richest qualitative anisotropy is apparent in the temperature range of $\sim$ 50~K, where $\rho_a$ and $\rho_b$ exhibit the classic drop in magnitude upon the onset of Kondo coherence, while $\rho_c$ begins to {\it increase} upon cooling, rising to a peak at 14 K before dropping precipitously. In the following, we compare the temperature dependence of resistivity with that of ARPES spectra, finding consistency with a Kondo lattice coherence picture for $a$-$b$ plane transport, and investigate magnetotransport and magnetization data to help elucidate the $c$-axis behavior.

Figure~(\ref{fig:ARPES}) presents an analysis of ARPES temperature dependence, with a representative spectrum along the $\Gamma - X$ axis shown in Fig.~(\ref{fig:ARPES})~(a). 
Integrating the region where dispersive bands are absent, we study the temperature dependence of the quasi-elastically scattered photoelectrons, as shown in  Fig.~(\ref{fig:ARPES})~(b). The peak within 0.1 eV of the Fermi level, which is cut by the resolution-convoluted Fermi function, follows a typical temperature evolution as coherence develops. As shown in the inset of Fig.~(\ref{fig:ARPES})~(b), tracking the peak magnitude as a function of temperature, an inflection can be seen around 50 K, where $\rho_a$ and $\rho_b$ rapidly drop. 
This is consistent with the formation of Kondo coherence near 50~K. 

To make further connection to transport, we focus on energies close to the Fermi energy (ideally $E-E_F \lessapprox k_{B}T$; however this energy window is not adequately resolved in the measurement so we use the closest available energy that can be analyzed). From the MDCs at 25 meV binding energy, we find the width of the Lorentzian fits (feature width) to be changing with temperature, as shown in Fig.~(\ref{fig:ARPES})~(d). Note that the fitting uncertainty is greater at higher temperatures due to irregular background intensity (see SM \cite{SM} Sect. VIII. for more detail). We can interpret that the temperature dependence of the ARPES feature width and the electrical resistivity is mainly governed by the temperature dependence of the mean free path of the carriers. 
The key finding is that the temperature evolution of the ARPES 6$d$ band feature width, as shown in Fig.~(\ref{fig:ARPES})~(d), is qualitatively consistent with the steadily decreasing behavior of $\rho_a$ and $\rho_b$ on cooling below the Kondo coherence temperature, and inconsistent with the rising behavior of $\rho_c$ in the same temperature range. Taken together with the behavior of the quasi-elastically scattered photoelectrons, this confirms the connection between the Kondo mechanism and $a$-$b$ plane resistivity and the anomalous distinction of $c$-axis transport.

Interestingly, the existence of an unusual qualitative anisotropy in resistivity temperature dependence has been observed in other systems
such as UCoGe \cite{UCoGe_Resistivity}, and is a well-known phenomenon in highly two-dimensional metals such as Sr$_2$RuO$_4$ \cite{NormalSr2RuO4} and cuprates, where its origin is still highly debated \cite{PRL_Gutman}.
In contrast to the two-dimensionally anisotropic systems, $c$-axis transport in UTe$_2$ is nearly equivalent in magnitude to its $b$-axis counterpart in this regime, suggesting other qualitative anisotropic scattering mechanisms must be at play. Further below, we discuss an analysis of magnetotransport and magnetization that suggests magnetism is responsible.

\begin{figure}[t]
\begin{center}%
\includegraphics[scale=1.2]{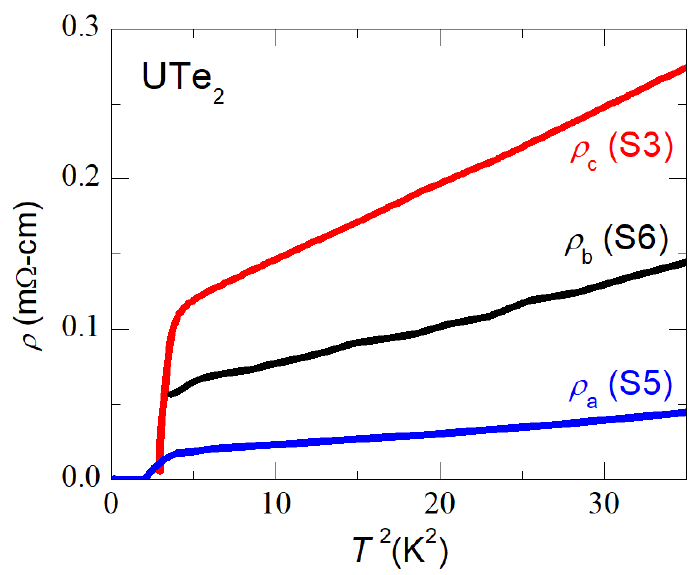}
\caption[]{(Color) Low-temperature resistivity of UTe$_2$, exhibiting Fermi liquid behavior for all three crystallographic orientations. Data were obtained from four-wire measurements on bar-shaped Samples (S3, S5, and S6).} 
\label{Fig:Fermiliquid_Kondo}
\end{center}
\end{figure}

\textit{Low Temperature} ($T_c< T <\sim$5 K):  
Upon cooling, it is not clear how the two-channel model discussed above evolves below the rich anisotropic features at intermediate temperatures, but all three resistivities indicate the realization of a heavy Fermi liquid-like state at low temperatures, decreasing substantially and approaching a saturating behavior with a  $T^2$ dependence as shown in Fig.~\ref{Fig:Fermiliquid_Kondo}.
(Note that bar-shaped samples are used for this analysis, using only sample data that agree with our generalized Montgomery technique measurements.)
This is surprising, in light of experimental evidence for strong spin fluctuations \cite{sundar2019coexistence} and quantum critical scaling \cite{ran2019nearly}, often associated with anomalous (i.e., non-Fermi liquid) scattering behavior. The $T^2$ coefficient ($A$), which is considered a measure of the strength of electron-electron interactions, is indeed enhanced in UTe$_2$ as expected from the moderately large electronic density of states observed in heat capacity \cite{ran2019nearly}, with values of 0.76, 2.56 and 5.03 $\mu\Omega$cm/K$^2$ for $\rho_a$, $\rho_b$ and $\rho_c$, respectively. 
The fact that all three coefficients are enhanced suggests that, however the band structure evolves through hybridization, all three conductivity components entail heavy band characteristics.
Furthermore, with the heaviest component along the $c$-axis (by a factor of 6.6 as compared with $\rho_a$), the anisotropy also evolves strongly as compared with a factor of $\sim$3 between the $c$- and $a$-axis resistivities at 20~K.
Lower-temperature ARPES experiments will help shed light on this evolution. 

\begin{figure*}[t]
    \centering
    \includegraphics[scale=0.8]{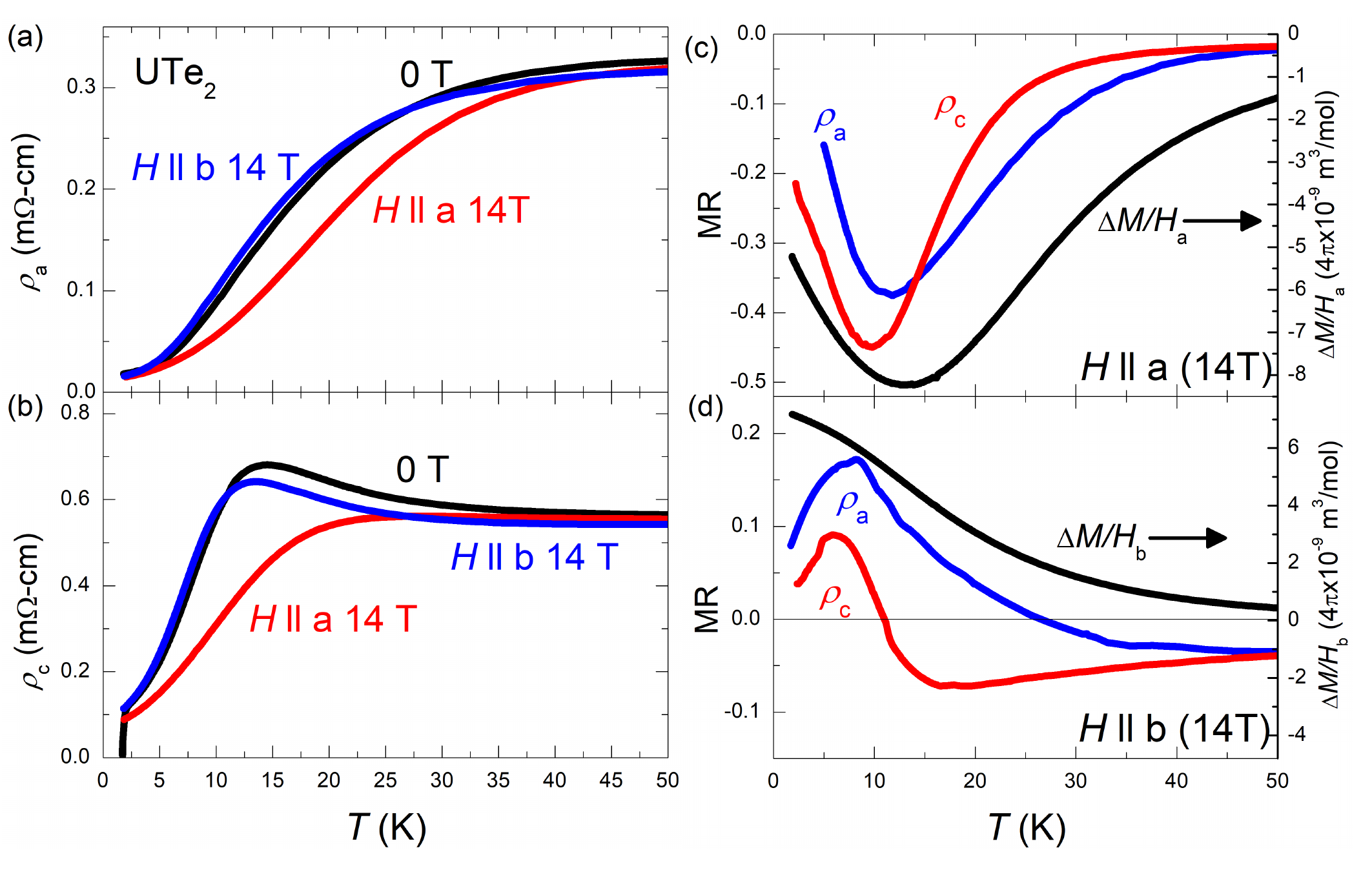}
    \caption{(Color) Magnetotransport results at 14T. (a) $\rho_{a}$ vs. temperature for both fixed at 0T (in black), $H \parallel a$ at 14T (in red), and $H \parallel b$ at 14T (in blue). Data were taken using bar-shaped sample S7. (b) $\rho_{c}$ vs. temperature for both fixed at 0T (in black), $H \parallel a$ at 14T (in red), and $H \parallel b$ at 14T (in blue). Data were taken using bar-shaped sample S3. (c) MR vs. temperature and $\Delta M_{a}/H_{a}$. Field is applied along the $a$-axis direction. (d) MR vs. temperature and $\Delta M/H_{b}$. Field is applied along the $b$-axis direction. Comparison of magnetoresistance for both $\rho_a$ and $\rho_c$ samples and $\Delta M/H$. Magnetoresistance (MR) is defined as MR$= (\rho(14 \textrm{T})-\rho(0 \textrm{T}))/\rho(0 \textrm{T})$ and $\Delta M/H = \chi_{CW} - M/H (\textrm{14 T})$, where $\chi_{CW}$ is the Curie-Weiss susceptibility fitted at high temperatures.  }
    \label{Fig:Resistivity_fixedField}
\end{figure*}

\textit{Magnetotransport:} 
To investigate the nature of the $c$-axis peak, we use field orientation dependent magnetoresistance (MR) as a probe of scattering anisotropy, focusing on whether MR exhibits a dependence on current or field direction.
In UTe$_2$, uranium atoms form chains along the magnetic easy axis ($a$-axis), with nearest neighbor ions forming ladder rungs parallel to the $c$-axis. 
We therefore compare the response of $\rho_a$ and $\rho_c$ MR with fields applied both parallel ($H\parallel a$) and perpendicular ($H\parallel b$) to the uranium chains (other field orientations are presented in the SM \cite{SM} Sect. VI.), expecting an anisotropic current response similar to the temperature dependence. 
Surprisingly, we find a nearly isotropic suppression of resistivity (i.e., negative MR) for both  $\rho_a$ and $\rho_c$ with fields applied along the magnetic easy axis ($a$-axis).
As shown in Fig.~(\ref{Fig:Resistivity_fixedField}), a negative MR is observed with $H\parallel a$ for both resistivities up to Kondo coherence, not only, notably, for the peak in $\rho_c$, but also for the broad inflection in $\rho_a(T)$. For $H\parallel b$, both $\rho_a$ and $\rho_c$ exhibit a small positive MR at the lowest temperatures with a crossover on warming. As shown in Figs.~(\ref{Fig:Resistivity_fixedField})~(c) and (d), the normalized MR shows this comparison more clearly, suggesting that the MR response does not depend heavily on the current direction, but rather mostly on the magnetic field orientation. Similar results have been obtained for UCoGe and ascribed to magnetic fluctuations \cite{Taupin}.  
Together with other reported observations, we take these results as evidence for the $c$-axis peak originating from a change in the magnetic spectrum.

An important reference is the magnetization at high fields. A Curie-Weiss (CW) susceptibility behavior, $M/H=\chi_{CW}$, was observed in UTe$_2$ at high temperatures for all three field orientations~\cite{ran2019nearly}, consistent with the behavior of a Kondo lattice system above its coherence temperature. However, at lower temperatures, deviations from CW behavior occur, with $M/H$ showing a maximum near 35~K for $H\parallel b$, and an inflection point near 10~K for $H\parallel a$ \cite{ran2019nearly}, with both features persisting to higher fields (see SM \cite{SM} Sect. V for all field orientations and different magnitudes). 
These features are comparable to those observed in our MR data. To emphasize this, we compare MR to the deviation of susceptibility from the CW behavior by plotting the difference ($\Delta M/H = \chi_{CW} - M/H$) for both $a$- and 
$b$-axis directions, shown in Fig.~(\ref{Fig:Resistivity_fixedField})~(c) and (d). 
We do this analysis for two reasons. First, this subtraction emphasizes the sub-leading order temperature dependence that only shows up as a mild slope change in the raw $M/H$ data. Second, the sign of $\Delta M/H$ indicates whether the susceptibility is changing faster or slower than the high-temperature CW behavior. For example, the CW behavior will saturate near the coherence temperature of a standard Kondo lattice, and therefore $\Delta M/H$ will be positive. For $H \parallel b$, we find that $\Delta M/H$ is indeed positive, but in contrast we find that $\Delta M/H$ is negative for $H \parallel a$. The maximum in $b$-axis magnetization (i.e., $\Delta M/H >0$) that occurs near the onset of Kondo coherence has been associated with an energy scale from the metamagnetic transition at 35 T \cite{miyake2019metamagnetic, miyake2021sharp}, while the inflection in $a$-axis magnetization near 10~K (i.e., $\Delta M/H < 0$) appears to be dominated by easy-axis magnetism of the uranium chains \cite{ran2019nearly}.
Interestingly, the comparison of MR and $\Delta M/H$ reveals a qualitative similarity in both the temperature trend and sign for both field orientations, especially the $\sim 10$~K negative peak feature for $H \parallel a$. This suggests that the change in scattering responsible for magnetotransport is predominantly magnetic in nature for both current directions.

Overall, the qualitative and quantitative differences observed between $a$-$b$ plane and $c$-axis transport, as well as the crossovers in resistivity anisotropy as a function of temperature, suggest that 1) at least two different transport channels are responsible for transport in different directions, and 2) the scattering mechanism(s) involves energy scales that are quite sensitive to the temperature range under study. In addition, from magnetotransport studies, 3) the peak in $\rho_{c}$ and minima in MR and $\Delta M/H$ for $H\parallel a$ occur at nearly the same temperature $\sim 10$~K, which is quite different from the Kondo coherence temperature
observed in $\rho_a$ and $\rho_b$ in Fig.~\ref{Fig:resistivityresults}. 
All of these observations can be explained by a scattering mechanism with a distinct $\sim$ 10~K energy scale that is magnetic (non-Kondo-like) in nature. 
For instance, this temperature is very close to the onset of quantum critical scaling of magnetization, with $M/T \propto H/T^{1.5}$ \cite{ran2019nearly}, suggesting that it coincides with a change in the fluctuation spectrum, while high temperature Curie-Weiss behavior indicates that antiferromagnetic interactions cannot be ignored. Details about the magnetic excitation spectrum are emerging \cite{PRL_Duan, knafo2021low,duan2021resonance, willa2021thermodynamic}, but may be challenging to interpret in a simple spin fluctuation picture due to the evolving heavy fermion band structure \cite{butch2022symmetry}.
Interestingly, nuclear magnetic resonance experiments \cite{tokunaga2019125te,tokunaga2022slow} have revealed a divergence in the spin–spin relaxation rate 1/$T_2$ only for $H\parallel a$, also suggesting the development of spin fluctuations below $\sim 20$~K and proximity to a (quasi) long-range ordered phase. In addition, given the absence of long-range magnetic order \cite{sundar2019coexistence}, the temperature scales observed in $\rho_c(T)$, the MR and the magnetic response suggest a magnetic crossover scale that dominates the $c$-axis transport channel.


This work provides a definitive measure of the electrical resistivity along all three primary axes of UTe$_{2}$ in the normal state. Given the expectation of strong anisotropy from electronic structure calculations, the magnitude of the $c$-axis resistivity is surprisingly comparable to that of the $a$- and $b$-axis resistivities in the entire temperature range, but exhibits a qualitative difference in behavior at temperatures below the onset of Kondo coherence. We understand this behavior as originating from electronic bands with distinct dimensionality, as well as a scattering mechanism that is intimately tied to a crossover in the magnetic spectrum near 15~K. Adding valuable information to our understanding of the normal state of UTe$_2$, this information will be important for understanding the electronic structure and for building a microscopic theory of superconductivity.  

\begin{acknowledgments}

We thank D. Agterberg, I. Hayes, Y. Furukawa, Q.P. Ding and C. Kurdak for inspiring discussions.
Research at the University of Maryland was supported by the Department of Energy Award No. DE-SC-0019154 (transport experiments), the National Science Foundation Division of Materials Research Award DMR-1905891 (support of J.C.), the Gordon and Betty Moore Foundation’s EPiQS Initiative through Grant No. GBMF9071 (materials synthesis), NIST, and the Maryland Quantum Materials Center. M.S.F. acknowledge support from the ARC Centre of Excellence in Future Low-Energy Electronics Technologies (FLEET, CE170100039). A. H. N. was supported by the National Science Foundation Division of Materials Research Award DMR-1917511 and by Robert A. Welch Foundation grant C-1818. This research used resources of the Advanced Light Source, a U.S. DOE Office of Science User Facility under Contract No. DE-AC02-05CH11231. Research at New York University was supported by the National Science Foundation under Grant No. DMR-2105081.

\end{acknowledgments} 

\nocite{dos2011procedure,miccoli2015100th,costi1994transport,zabrodskii2001coulomb,Shirley,lee2011electrolyte,YSEoInverted,andrei1983solution,parks2010mechanical}

\bibliography{bibliography}

\end{document}